

\input harvmac
\def\listrefssave
{\footatend\bigskip\bigskip\bigskip\immediate\closeout\rfile\writestoppt
\baselineskip=14pt\centerline{{\bf References}}\bigskip{\frenchspacing%
\parindent=20pt\escapechar=` \input \jobname.refs\vfill\eject}\nonfrenchspacing}
\def\Titleh#1#2{\nopagenumbers\abstractfont\hsize=\hstitle\rightline{#1}%
\vskip .5in\centerline{\titlefont #2}\abstractfont\vskip .5in\pageno=0}
\def\CTPa{\it Center for Theoretical Physics, Department of Physics,
      Texas A\&M University}
\def\CTPb{\it College Station, TX 77843-4242, USA}
\def\HARCa{\it Astroparticle Physics Group,
Houston Advanced Research Center (HARC)}
\def\HARCb{\it The Woodlands, TX 77381, USA}
\def\CERN{\it Theory Division, CERN, 1211 Geneva 23, Switzerland}

\def\nextline{\unskip\nobreak\hfill\break}

\catcode`\@=11 

\def\lsim{\mathrel{\mathpalette\@versim<}}
\def\gsim{\mathrel{\mathpalette\@versim>}}
\def\@versim#1#2{\vcenter{\offinterlineskip
    \ialign{$\m@th#1\hfil##\hfil$\crcr#2\crcr\sim\crcr } }}
\def\boxit#1{\vbox{\hrule\hbox{\vrule\kern3pt
      \vbox{\kern3pt#1\kern3pt}\kern3pt\vrule}\hrule}}

\def\t1{{\tilde 1}}

\def\DVN{D. V. Nanopoulos}

\def\GeV{\,{\rm GeV}}
\def\TeV{\,{\rm TeV}}

\def\st{\sin^2\theta_W(m_{Z})}
\def\ae{\alpha_{em}(m_{Z})}
\def\as{\alpha_3(m_Z)}

\def\NPB#1#2#3{Nucl. Phys. B {\bf#1} (19#2) #3}
\def\PLB#1#2#3{Phys. Lett. B {\bf#1} (19#2) #3}

\def\PRD#1#2#3{Phys. Rev. D {\bf#1} (19#2) #3}
\def\PRL#1#2#3{Phys. Rev. Lett. {\bf#1} (19#2) #3}

\nref\GeorgiI{H. Georgi and S.L. Glashow, \PRL{32}{74}{438}.}
\nref\GeorgiII{H. Georgi, H. Quinn and S. Weinberg, \PRL{33}{74}{451}.}
\nref\EKNI{J. Ellis, S. Kelley, and \DVN, \PLB{249}{90}{441}.}
\nref\AMALDI{U. Amaldi, W. de Boer and H. Furstenau, \PLB{260}{91}{447}.}
\nref\EKNII{J. Ellis, S. Kelley and \DVN, \PLB{260}{91}{131}.}
\nref\SU{S. Dimopoulos and H. Georgi, \NPB{193}{81}{150}; \nextline
N. Sakai, Z. Phys. {\bf{C11}} (1982) 153;\nextline
A. Chamseddine, R. Arnowitt and P. Nath, \PRL{105}{82}{970}.}
\nref\EKNIII{J. Ellis, S. Kelley and \DVN, CERN preprint CERN-TH.6140/91
(1991), to be published in Nucl. Phys. B.}
\nref\ACPZ{F. Anselmo, L. Cifarelli, A. Petermann and A. Zichichi,
CERN-TH.6429/92 (1992).}
\nref\EF{J. Ellis, G. Fogli and E. Lisi, \PLB{274}{92}{456} and
references therein.}
\nref\DEGRASSI{G. Degrassi, S. Fanchiotti and A. Sirlin,
\NPB{351}{91}{49}.}
\nref\HADRONIC{H. Burkhardt, F. Jegerlehner, G. Penso and
C. Verzegnassi,
Z. Phys. {\bf{C11}} (1989) 497.}
\nref\ASTRONG{ALEPH Collaboration, D. Decamp et al.,
CERN Preprint PPE/92-33 (1992); \nextline
DELPHI Collaboration, P. Abreu et al., CERN Preprint
PPE/91-181/Rev. (1992);\nextline
L3 Collaboration, O. Adriani et al., CERN Preprint
PPE/92-58 (1992);\nextline
OPAL Collaboration, P.D. Acton et al., CERN Preprint
PPE/92-18 (1992);\nextline
A.X. El Khadra, G. Hockney, A.S. Kronfeld and
P.B. Mackenzie, Fermilab Preprint PUB-91/354-T (1991);\nextline
W. Kwong, P.B. Mackenzie, R. Rosenfeld and J.L. Rosner,
\PRD{37}{88}{3210};\nextline
A.D. Martin, R.G. Roberts and W.J. Stirling, Durham Preprint
DTP 90-76 (1990), Rutherford Preprint RAL-91-044 (1991).}
\nref\ROSS{J. Ellis, D.V. Nanopoulos and D. Ross, \PLB{267}{91}{132}.}
\nref\PDECAY{B. Campbell, J. Ellis and D.V. Nanopoulos,
\PLB{141}{84}{229}; \nextline
R. Arnowitt and P. Nath, \PRD{38}{88}{1479}.}
\nref\ZUT{H. Dreiner, J.L. Lopez, D.V. Nanopoulos and
D. Reiss, \PLB{216}{89}{289}.}
\nref\FLIP{I. Antoniadis, J. Ellis, J. Hagelin, and D.V. Nanopoulos,
\PLB{194}{87}{231};
{\bf B205} (1988) 459;  {\bf B208} (1988) 209;
{\bf B231} (1989) 65.}
\nref\EUX{I. Antoniadis, J. Ellis, R. Lacaze and D.V. Nanopoulos,
\PLB{268}{91}{188} and references therein.}
\nref\SISM{S. Kelley, J. Lopez and D.V. Nanopoulos, \PLB{278}{92}{140}.}

\Titleh{\vbox{\baselineskip12pt
\hbox{CERN--TH.6481/92}
\hbox{CTP--TAMU--42/92}
\hbox{ACT--10/92}}}
{\vbox{\centerline{
Constraints From Gauge Coupling Unification}
\vskip2pt\centerline{
On The Scale Of Supersymmetry Breaking}}}
\centerline{\bf John Ellis}
\bigskip
\centerline{\CERN}
\bigskip
\centerline{{\bf S. Kelley and D.~V.~Nanopoulos}\foot{\it Present
address: Theory Division, CERN, 1211 Geneva 23, Switzerland.}}
\bigskip
\centerline{\CTPa}
\centerline{\CTPb}
\centerline{and}
\centerline{\HARCa}
\centerline{\HARCb}
\vskip .3in
\centerline{ABSTRACT}
We reanalyze precision LEP data and coupling
constant unification in the minimal supersymmetric
$SU(5)$ model including the evolution of the
gaugino masses.  We derive general
bounds on the primordial gaugino
supersymmetry-breaking mass-scale
$m_{1/2}$ in terms of the
 various input parameters.
The model cannot accommodate $m_{1/2}<1\TeV$ for values of
$\as < 0.115$, even for extreme $1-\sigma$ values of the other
inputs.
We emphasize the sensitivity of this type of
calculations to the various input parameters.
\bigskip
\leftline{CERN-TH.6481/92}
\leftline{CTP-TAMU-42/92}
\leftline{ACT-10/92}
\leftline{April 1992}
\Date{}

\newsec{Introduction}
Gauge coupling unification has always led to attractive predictions of
Grand Unified Theories, GUTs \GeorgiI. Although qualitatively successful,
minimal non-supersymmetric
GUTs predict $\st$ about 8 to 5 percent lower than the present LEP
value \GeorgiII. However, minimal supersymmetric GUTs predict
$\st$ tantalizingly
close to the LEP value \EKNI.  Some authors have suggested that the
amazingly close agreement of LEP data with minimal supersymmetric GUTs
is evidence for supersymmetry at a scale near $1\TeV$ \AMALDI, in accord
with theoretical prejudice based on naturalness arguments.
  Previously, we
developed analytically one-loop expressions including new particle
 thresholds, and
corrected these expressions to two-loop accuracy \EKNII.  These expressions
were then applied to to a specific model, minimal supersymmetric
$SU(5)$ \SU ,
in an attempt to bound the supersymmetry-breaking scale in the form of a
 universal primordial
gaugino mass, $m_{1/2}$ \EKNIII.

We found that the data then available favored $m_{1/2}<65\GeV$ or
$m_{1/2}>21\TeV$ at the one-standard-deviation ($1-\sigma$) level.
However, in those papers we did not include the
dependences of the physical gaugino mass thresholds on the
gauge couplings, which has recently been shown to have a significant
impact on the favored range of $m_{1/2}$: the evolution of gaugino
masses (EGM) effect \ACPZ.

In this letter, we correct and
continue our previous work \EKNIII\ on the derivation of
rigorous bounds on $m_{1/2}$, incorporating this EGM effect and the
latest available LEP data.  Since the largest uncertainty in $m_{1/2}$
is now that due to $\as$, we present our results as bounds on $m_{1/2}$
as a function of $\as$ for specific values and plausible
ranges of the other
parameters: $\st$, $\ae$, $m_t$, the Higgs superpotential
 mixing parameter
$\mu$, and the Higgs mass $m_h$.  Throughout this
paper, $\st$ and all gauge couplings are defined using the
$\overline{MS}$ scheme.
Using extreme $1-\sigma$ values of the other inputs, we
find that $m_{1/2}<1\TeV$ can be obtained only for $\as>
 0.115$.
We show how to modify our results for different values of the inputs,
noting that improved precision will
most likely increase the minimum value of $\as$ compatible
with $m_{1/2}<1\TeV$.

\newsec{Basic Formulae and LEP data}

The prediction for $\st$ in minimal supersymmetric $SU(5)$ may be
written as \EKNIII
\eqn\I{\st=0.2+{{7\ae}\over{15\as}}+0.0029+\delta_s(light)
+\delta_s(heavy)+\delta_s(conv)}
where 0.0029 corrects the analytic one-loop calculation to two-loop
accuracy.  The scheme conversion term $\delta_s(conv)$ is negligible.
Because of constraints on the Higgs triplets mediating proton decay,
the heavy particle threshold correction
$\delta_s(heavy)>0$ in minimal supersymmetric $SU(5)$ \EKNIII.
The contribution from light particle thresholds is:
\eqna\II
$$\eqalignno{\delta_s(light)=&{{\ae}\over{20\pi}}
[-3ln({{m_t}\over{m_Z}})
+{28\over 3}ln({c_{\tilde g}m_{1/2}\over m_Z})
-{32\over 3}ln({c_{\tilde w}m_{1/2}\over m_Z})\cr
&-ln({{m_h}\over{m_Z}})-4ln({{\mu}\over{m_Z}})+{4\over 3}f(m_{1/2},y,w)]
&\II {}\cr}$$
where
\eqna\III
$$\eqalignno{
f(m_{1/2},y,w)=&{{15}\over 8}ln({{m_{1/2}\sqrt{c_{\tilde q}+y}}\over{m_Z}})
-{9\over 4}ln({{m_{1/2}\sqrt{c_{\tilde e_l}+y}}\over{m_Z}})\cr
&+{3\over 2}ln({{m_{1/2}\sqrt{c_{\tilde e_r}+y}}\over{m_Z}})
-{{19}\over{48}}ln({{m_{1/2}\sqrt{c_{\tilde q}+y+w}}\over{m_Z}})\cr
&-{{35}\over{48}}ln({{m_{1/2}\sqrt{c_{\tilde q}+y-w}}\over{m_Z}})
&\III {}\cr}$$
and $y \equiv (m_{0}/m_{1/2})^2$, where $m_{0}$ is a universal primordial
 supersymmetry-breaking spin-zero mass, $w$ was defined in \EKNIII\
and the logarithms should be set to zero if the threshold is below $m_Z$.

The value of $\st$ is known from an analysis of precision data from
LEP and elsewhere:
\eqn\IV{\st=0.2327\pm 0.0007 \quad \EF}
The value of $\ae$ is also well-known:
\eqn\V{\ae={1\over{127.9\pm 0.2}}\quad  \DEGRASSI\HADRONIC}
However, a glance at Table
1 shows that there is a huge range in the experimental determinations of
$\as$ \ASTRONG.  Naively taking an equally-weighted mean of these eight
different values and the uncertainty of the least uncertain experiment
gives $\alpha_3(m_Z)=.116\pm .005$.
However, it is disturbing that the different experimental determinations
of $\alpha_3(m_Z)$ do not overlap.  Except for the DELPHI value, all the
LEP measurements are higher than, and do not even overlap the other three
lower-energy
measurements.  It is known that higher-order corrections to these LEP
values are important, and preliminary attempts to sum the dominant
contributions to the perturbative higher-order effects indicate that
the LEP results could be shifted down to values nearer
the other measurements \ROSS.

We take an agnostic
approach in this paper and derive bounds on $m_{1/2}$
for arbitrary values of $\alpha_3(m_Z)$, and await further
progress in its determination.

\newsec{Derivation of Bounds on $m_{1/2}$}

In this section, we use the information of the previous section to
derive bounds on $m_{1/2}$ for different values of $\as$,
assuming $\delta_s(heavy)=0$\foot{If $\delta_s(heavy)>0$ the bounds will be
even
tighter.} and ignoring the negligible $\delta_s(conv)$.

Consider first a scenario where the winos are lighter than $m_Z$
but the gluinos are heavier than $m_Z$.
The formula \I\ then gives:
\eqn\VI{
ln({{m_{1/2}}\over{m_Z}})={1\over 7}X
-{\pi\over{\alpha_3}}-c_{\tilde g}}
where
\eqna\VII
$$\eqalignno{
X=&{{15\pi}\over{\ae}}(\st-.2029)\cr
&+{9\over 4}ln({{m_t}\over{m_Z}})+3ln({{\mu}\over{m_Z}})
-{3\over 4}ln({{m_h}\over{m_Z}})
-f(m_{1/2},y,w)&\VII {}\cr}$$
Now consider the case where both the gluinos and the winos are heavier
than $m_Z$, in which case \I\  gives:
\eqn\VIII{
ln({{m_{1/2}}\over{m_Z}})=
-X+{{7\pi}\over{\alpha_3}}+7c_{\tilde g}-8c_{\tilde w}}
To obtain the most generous bounds from \VI\  and \VIII, we would need to
maximize $X$, minimize $c_{\tilde g}$, and
maximize $c_{\tilde w}$.

For physically-relevant values of $w$ (those which give positive squared masses
for the stop squarks), $f(m_{1/2},y,w)$
is minimized at $w=-8(c_{\tilde q}+y)/27$.
With this value of $w$, $f(y,w)$ has one extremum, a maximum, at
$y=(c_{\tilde l_l}c_{\tilde l_r}+2c_{\tilde l_l}c_{\tilde q}
-c_{\tilde l_r}c_{\tilde q})/(3c_{\tilde l_l}-2c_{\tilde l_r}-c_{\tilde q})$,
and approaches -0.025 as y becomes very large.  Since the values of the
$c's$ that we encounter satisfy $c_{\tilde q}>1>c_{\tilde l_l},c_{\tilde l_r}$,
the minimum of $f(y,w)$ is indeed -0.025 for values of $y>0$.
We have searched numerically the region where some of the
scalars contibuting to $f(m_{1/2},y,w)$ are lighter than $m_Z$ to verify
that it does not take smaller values in this region.

Taking the extreme $1-\sigma$ values $\st=0.2334$, $m_t=95\GeV$,
$\ae=1/128.1$ and using a
naturalness bound of $1\TeV$ on $\mu$ and $m_h$ gives a
maximum numerical value for $X$ of
193.2.  Note that the
extreme values of $\st$ and $m_t$ are correlated, and that the contribution
of the squarks and sleptons which is represented by $f(y,w)$ is
negligible.

Estimating the values of $c_{\tilde g}$ and $c_{\tilde w}$ is more
involved, since they depend on the unkown value of $m_{1/2}$ and an
iterative procedure must be used \ACPZ.
At the one-loop
\eqn\IX{
c_{\tilde g}={{\alpha_3(c_{\tilde g}m_{1/2})\over{\alpha_U}}}\quad
c_{\tilde w}={{\alpha_2(c_{\tilde w}m_{1/2})\over{\alpha_U}}}}
where $c_{\tilde g}m_{1/2}$ and $c_{\tilde w}m_{1/2}$ are the gluino
and wino masses renormalized at $m_{\tilde g},m_{\tilde w}$.

 From the one-loop expression for renormalizing a coupling
from $m_Z$ up to its gaugino threshold:
\eqn\X{
\alpha(m_{gaugino})={{\alpha(m_Z)}\over{1-{b\over{2\pi}}
ln({{\alpha(m_{gaugino})m_{1/2}}\over{m_Z\alpha_U}})}}}
We see that, for $b<0$, $\alpha(m_{gaugino})$
increases with $b$.  In order to minimize $c_{\tilde g}$, we want
to use the minimum value of $b_3=-7$ possible in the MSSM below the
gluino threshold.  Similarly, to maximize
$c_{\tilde w}$ we use the maximum value of $b_2=-1/3$ possible
below the wino threshold.

Fitting the results of an analytic
 one-loop calculation to a numeric two-loop
calculations for central values gives \EKNIII:
\eqna\XI
$$\eqalignno{{1\over{\alpha_U}}=
&{3\over{20\ae}}+{1\over{5\alpha_3(m_Z)}}-0.7
+{1\over{5\pi}}[3ln({{m_{\tilde g}}\over{m_Z}})
+{1\over 8}ln({{m_{h}}\over{m_Z}})
+{3\over 8}ln({{m_{\tilde l_l}}\over{m_Z}})\cr
&+{3\over 8}ln({{m_{\tilde l_r}}\over{m_Z}})
+{{17\over 4}}ln({{m_{\tilde q}}\over{m_Z}})
+{{83}\over{48}}ln({{m_t}\over{m_Z}})
+{1\over 2}ln({{m_{\tilde w}}\over{m_Z}})
+{1\over 2}ln({{\mu}\over{m_Z}})]&\XI {}\cr}$$
where the stop squarks have been taken degenerate with the other squarks.
Thus, $\alpha_U$ decreases with the thresholds.  Taking upper bounds of
$165\GeV$ on the top mass [9]
and $3\TeV$ on the other thresholds gives the range
\eqn\XII{{3\over{20\alpha_{em}}}+{1\over{5\alpha_3(m_Z)}}-0.7<
{1\over{\alpha_U}}<
{3\over{20\alpha_{em}}}+{1\over{5\alpha_3(m_Z)}}+1.4}
for the coupling at the unification scale.
Numerically, we find a slight variation of the
solutions of \VI\ and \VIII\ over
this range of $\alpha_U$, with both values increasing with $\alpha_U$.
Therefore, we use the maximum value in \VI\ and the minimum value in \VIII.

Finally, we use the central
value of $\ae$ to calculate $c_{\tilde w}$ and $\alpha_U$,
neglecting the small variations from the uncertainty in $\ae$.

\newsec{Results and Conclusions}

Fig. 1 shows on the left
the triangular region excluded by the calculation of
the previous
section, which is based on the formulae of Ref. [7]
with the EGM effect incorporated as suggested in Ref. [8].
The upper solid line is
obtained by iterating \VIII, describing a scenario
where $m_{\tilde w}>m_Z$, and the lower solid line is obtained
by iterating \VI, describing a scenario where $m_{\tilde w}<m_Z$.  The area
between these two solid lines is excluded.  An
approximate lower bound of $m_{1/2}>45\GeV$ from CDF gluino searches and
an upper naturalness bound $m_{1/2}<1\TeV$ are represented as dashed
lines.  To have $m_{1/2}<1\TeV$, $\as>0.115$ is needed.
Note that the $J/\Psi$, $\Upsilon$ and deep inelastic values of $\as$
are below this lower bound! If the lower
bound on $m_{1/2}$ can be pushed to $60\GeV$, $m_{1/2}>1\TeV$ will then
require $\as>0.116$.

Table 2 shows the value of $X$ for different values of the inputs within
their $1-\sigma$ range.
Since our results use extreme $1-\sigma$ values to
maximize $X$, $X$ can only decrease if these errors are reduced
, giving an even tighter lower bound on
$\as$.  For reference, we also show in fig. 1 the bounds from
$X=189.4$ corresponding to central input values as a dashed line, and
the minimum value of $X=185.4$ as a dotted line.  Note these bounds are
very sensitive to the values of the input parameters.  Since the various
approximations which have been made all loosen the bound on $\as$, the
bounds would become stricter if our hypotheses could be tightened.
Whilst the minimal supersymmetric $SU(5)$  GUT
model cannot be ruled out on the basis of coupling
constant unification, things are becoming very tight.  Explaining the
discrepancy between the different determinations of $\as$ is crucial.
If one believes the high values,
or even the average values of $\as$, the model
is in good shape.  However, if higher-order corrections [13]
do actually bring
the high values of $\as$ down to
match the low values, then coupling constant
unification cannot be achieved in the minimal model at
the $1-\sigma$ level.

We conclude this paper with a brief explanation of our attitude towards
these detailed estimates of the supersymmetry-breaking scale in the
minimal supersymmetric $SU(5)$ model. We regard these calculations as
illustrative, relevant for developing the technology needed to calculate
this scale in any explicit model, and indicating how far one can get on
the basis of our present experimental knowledge using the theoretical
tools available, but we do not take the minimal supersymmetric $SU(5)$
model as seriously as do some of our colleagues. Even in a purely
field-theoretical context, this model has to be modified if the Higgs
doublets and triplets are to be split in a natural way, and possible
difficulties with too-rapid proton decay are to be avoided \PDECAY. More
fundamentally, one cannot nowadays be blind to the promise and
successes of string theory in unifying quantum gravity with the other
particle interactions. It is known \ZUT\ that
the adjoint Higgs multiplet needed in
the minimal supersymmetric $SU(5)$ model cannot be obtained from string
theories in which gauge charges appear at the Kac-Moody level 1, which
is the formulation used in all models constructed so far. Among existing
string models, the one closest in philosophy to traditional $SU(5)$ is
the flipped $SU(5)$ model \FLIP, which also
incorporates natural Higgs doublet-
triplet splitting. This model offers the possibility that $\st$ could be
smaller than in minimal supersymmetric $SU(5)$, which is not
inconsistent with the data, particularly if the lower quoted values of
$\as$ turn out to be correct. In this and many
other string-derived models
the value of the string unification scale at which all the gauge
couplings appear to become equal is calculated \EUX\
to be significantly
higher than the grand unification scale calculated in minimal
supersymmetric $SU(5)$. Within the context of flipped $SU(5)$, this
reinforces the suggestion that $\st$ might be decreased. More
generally, this scale discrepancy suggests the existence of additional
particles beyond those in the Standard Model or its minimal GUT
extension. One example, the String Inspired Standard Model (SISM), was
proposed in ref. \SISM. It would be desirable to carry out in this model
and others derived from string calculations as detailed as those
described in this and our previous papers, but such a study would take
us beyond the scope of this note. The reader should keep in mind,
though, the observation that analyses of precision LEP data can be
used to test string ideas, as well as the GUT ideas discussed in the
bulk of this paper. Yet more motivation, as if our experimental
colleagues needed any, to carry on compressing the LEP error bars,
in particular for $\as$.
\vskip 0.5cm
\noindent
{\bf Acknowledgements}

The work of D.V.N. is partially supported by DOE
grant DE-FG05-91-ER-40633.

\vskip 0.5cm
\noindent
{\bf Figure Caption}

The allowed ranges of $m_{1/2}$ for different values of $\as$ are shown,
in the context of minimal supersymmetric $SU(5)$.
  The value of $m_{1/2}$ for the central values of the other inputs is
displayed as a dashed line. The minimal value of $\as$
assuming $1-\sigma$
variations of the other inputs is shown as a solid line, and the
maximal $1-\sigma$ value of $\as$ as a dotted line. For any given value
of $\as$, values of $m_{1/2}$ to the left of the solid line, or to the
right of the dotted line, are accordingly disfavoured. Also shown as
horizontal dashed lines are a lower bound $m_{1/2} >$ 45 GeV from CDF
gluino searches, and an upper bound $m_{1/2} <$ 1 TeV motivated by
naturalness. It can be seen that $m_{1/2} <$ 1 TeV is favoured only if
$\as >$ 0.115.

\vfill
\break

\input tables
{\hfill
{\begintable
\ Experiment \ \|\ Central Value \ \|\ Error \crthick
ALEPH  jets \|\ ~~$0.125$ \| ~~$\pm 0.005$ \nr
DELPHI jets \|\ ~~$0.113$ \| ~~$\pm 0.007$ \nr
L3 jets \|\     ~~$0.125$ \| ~~$\pm 0.009$\nr
OPAL   jets \|\ ~~$0.122$ \| ~~$\pm 0.006$ \nr
OPAL   $\tau$ \|\ ~~$0.123$ \| ~~$\pm 0.007$ \nr
$J/\Psi$ \|\ ~~$0.108$ \| ~~$\pm 0.005$ \nr
$\Upsilon$ \|\ ~~$0.109$ \| ~~$\pm 0.005$ \nr
Deep Inelastic \|\ ~~$0.109$ \| ~~$\pm 0.005$ \nr
Average \|\ ~~$0.116$ \| ~~$\pm 0.005$ \endtable}
\hfill}
\smallskip
\parindent=0pt
\centerline{
{\it Table 1} - Experimental Values of $\alpha_3(m_Z)$.}

\vskip 2cm

{\hfill
{\begintable
\ Input Values \ \|\ $\st$ \ \|\ $m_t$ \ \|\ $\ae$ \|\ $m_h$ \|\ $\mu$ \|\ $X$
 \crthick
$X_{max}$
\|\ ~~$0.2334$ \| ~~$95\GeV$ \| ~~$1/128.1$ \|
 $1\TeV$ \| $1\TeV$ \| $193.2$ \nr
central $\st$,$m_t$
\|\ ~~$0.2327$ \| ~~$130\GeV$ \|
 ~~$1/128.1$ \| $1\TeV$ \| $1\TeV$ \| $189.7$ \nr
central $\ae$
\|\ ~~$0.2334$ \| ~~$95\GeV$ \| ~~$1/127.9$ \|
 $1\TeV$ \| $1\TeV$ \| $192.9$ \nr
$X_{central}$
\|\ ~~$0.2327$ \| ~~$130\GeV$ \| ~~$1/127.9$ \|
 $1\TeV$ \| $1\TeV$ \| $189.4$ \nr
$X_{min}$
\|\ ~~$0.2320$ \| ~~$160\GeV$ \| ~~$1/127.7$ \|
 $1\TeV$ \| $1\TeV$ \| $185.4$ \nr
$X_{max}$: $\mu,m_h<500\GeV$
\|\ ~~$0.2334$ \| ~~$95\GeV$ \| ~~$1/128.1$ \|
 $500\GeV$ \| $500\GeV$ \| $190.6$ \endtable}

\hfill}
\smallskip
\parindent=0pt
\centerline{
{\it Table 2} - Sensitivity of X to various inputs:}
\vskip1pt\centerline{$\mu ,m_h < 1$ TeV unless otherwise stated.}

\vfill
\break

\listrefssave
\bye